\documentclass[prd,onecolumn,showpacs,floatfix,superscriptaddress,nofootinbib]{revtex4-2}
\usepackage{graphicx}
\usepackage{epsfig}
\usepackage{bm}
\usepackage{amssymb}
\usepackage{float}
\usepackage{amsmath}
\usepackage{dcolumn}
\usepackage{cancel}

\usepackage{mathrsfs}
\usepackage{dcolumn}
\usepackage{graphicx}
\usepackage{amsmath}
\usepackage{amsfonts}
\usepackage{amssymb}
\usepackage{microtype}
\usepackage{subfigure}
\usepackage{makeidx}
\usepackage{bm}
\usepackage{epsf}
\usepackage{color}
\usepackage{multirow,dcolumn}
\usepackage{graphicx}
\usepackage{mathrsfs}
\graphicspath{{Images/}}

\def\doi{http://doi.org}

\def\be{\begin{equation*}}
\def\ee{\end{equation*}}

\def\Ref{\ref}

\begin{document}

\title{Black holes of $(2+1)$-dimensional $f(R)$ gravity coupled to a scalar field}

\author{Thanasis Karakasis}
\email{asis96kar@gmail.com} \affiliation{Physics Division,
National Technical University of Athens, 15780 Zografou Campus,
Athens, Greece.}

\author{Eleftherios Papantonopoulos}
\email{lpapa@central.ntua.gr} \affiliation{Physics Division,
National Technical University of Athens, 15780 Zografou Campus,
Athens, Greece.}

\author{Zi-Yu Tang}
\email{tangziyu@sjtu.edu.cn}
\affiliation{Center for Astronomy and Astrophysics, School of Physics and Astronomy,
Shanghai Jiao Tong University, Shanghai 200240, China}

\author{Bin Wang}
\email{wang\_b@sjtu.edu.cn}
\affiliation{School of Aeronautics and Astronautics, Shanghai Jiao Tong University, Shanghai 200240, China}
\affiliation{Center for Gravitation and Cosmology, College of Physical Science
and Technology, Yangzhou University, Yangzhou 225009, China}

\vspace{17.5cm}

\begin{abstract}

 We consider $f(R)$ gravity theories in the presence of a scalar field minimally coupled to gravity with a self-interacting potential in $(2+1)$-dimensions. Without specifying the form of the $f(R)$ function, we first obtain an exact black hole solution dressed with  scalar hair  with the scalar charge to appear in the $f(R)$ function and  we  discuss its thermodynamics. This solution at large distances gives a hairy BTZ
black hole, and it reduces to the BTZ black hole when the scalar field decouples. In a pure $f(R)$ gravity supported by
the scalar field, we find an exact hairy black hole similar to the BTZ black hole with phantom hair and an analytic $f(R)$ form and discuss its thermodynamics.

\end{abstract}

\maketitle

\flushbottom

\tableofcontents

\section{Introduction}

 In three-dimensions one of the first exact black holes with a negative cosmological constant was discussed by Ba\~{n}ados,  Teitelboim, and Zanelli (BTZ) \cite{BTZ1,BTZ2}. This solution  came as a surprise in the scientific community, since in three dimensions the Weyl tensor, describing the distortion of the shape of a body in the presence of tidal  gravitational force, vanishes by definition while the Ricci tensor, describing  how the volume of the body changes due to this tital force, vanishes in the absence of matter. Therefore, since $Riemann = Weyl + Ricci$ we can only have flat spacetime. In this solution, the existence of a cosmological constant term and the electromagnetic field results to a non-zero Ricci tensor allowing in this way the existence of a black hole solution.

 After the discovery of this solution, scalar fields minimally and nonminimally coupled to gravity were introduced as matter fields. In \cite{Martinez:1996gn,Henneaux:2002wm} three dimensional black holes with a conformally coupled scalar field, being regular everywhere, were discussed.  After these first results other hairy black holes in three-dimensions were discussed  \cite{CMT1,CMT2,Correa:2012rc,Natsuume,Aparicio:2012yq,Xu:2014uha}. In \cite{Cardenas:2014kaa}  three-dimensional gravity with negative cosmological constant in the presence of a scalar field and an Abelian gauge field was introduced. Both fields are conformally coupled to gravity, the scalar field through a nonminimal coupling with the curvature and the gauge field by means of a Lagrangian given by a power of the Maxwell one. A sixth-power self-interaction potential, which does not spoil conformal invariance is also included in the action, resulting to a very simple relation between the scalar curvature and the cosmological constant. In \cite{Xu:2014uka} and \cite{Xu:2013nia} $(2+1)$ dimensional charged black holes with scalar hair where derived, where the scalar potential is not fixed ad hoc but instead derived from Einstein's equations. In \cite{Tang:2019jkn} exact three dimensional black holes with non-minimal scalar field were discussed. Finally, in \cite{Chan:1994qa} and \cite{Chan:1996rd}, static black holes in three dimensional dilaton gravity and modifications of the BTZ black hole by a dilaton/scalar were investigated.

 In four dimensions, the first black hole solution with a scalar field as a matter field was derived by Bocharova, Bronnikov and  Melnikov and independently by Bekenstein, called BBMB black hole \cite{BBMB}. The scalar field is conformally coupled to gravity, resulting to the vanishing of scalar curvature, the metric takes the form of the extremal Reissner-Nordstrom spacetime and the scalar field diverges at the horizon. It was also shown at \cite{bronnikov} that this solution is unstable under scalar perturbations. Later, a scale was introduced to the theory via a cosmological constant at \cite{Martinez:2002ru} and also a quartic scalar potential that does not break the conformal invariance of the action, resulting at a very simple relation between the scalar curvature and the cosmological constant. The scalar field does not diverge at the horizon, but the solution is found to be unstable \cite{Harper:2003wt}. In \cite{Anabalon:2012tu} asymptotically (anti) de Sitter black holes and wormholes with a self interacting scalar field in four dimensions were discussed. Regarding the minimal coupling case, the first exact black hole solution was presented at \cite{Martinez:2004nb}, the MTZ black hole. The scalar potential is fixed ad hoc, the geometry of the solution is hyperbolic and the scalar field remains finite at the black hole horizons. In \cite{Martinez:2006an} the electrically charged case is discussed. In \cite{Kolyvaris:2009pc}, a potential that breaks the conformal invariance of the action of the MTZ black hole in the Jordan frame was considered and new black hole solutions where derived. In \cite{Gonzalez:2013aca} the scalar field is fixed ad hoc and novel black hole solutions are investigated, letting the scalar potential to be determined from the equations and in \cite{Gonzalez:2014tga} the electrically charged case is considered. In \cite{Rinaldi:2012vy, Anabalon:2013oea} black holes with nonminimal derivative coupling were studied.  However, the scalar field which was coupled to Einstein tensor should be considered as a particle living outside the horizon of the black hole because it blows up on the horizon. Finally, in \cite{Cisterna:2021xxq} Plebanski-Demianski solutions in Quadratic gravity with conformally coupled scalar fields were investigated.

  The $f(R)$ theories of gravity were mainly introduced in an attempt to describe the early and late cosmological history of our Universe  \cite{DeFelice:2010aj}-\cite{Starobinsky:1980te}.  In particular, following the  recent cosmological observational results the $f(R)$ gravity cosmological models were used to explain the deceleration-acceleration transition. This requirement imposed constraints on  the $ f(R)$ models allowing viable choices of $ f(R)$ \cite{Capozziello:2014zda}.
   These theories  exclude contributions from any curvature invariants other than $R$ and they avoid the Ostrogradski instability \cite{Ostrogradsky:1850fid} which usually is present in higher derivative theories \cite{Woodard:2006nt}.  Several black hole solutions in these theories were found  and they either are deviations of the known black hole solutions of General Relativity, or they possess new properties that should be investigated. Static and spherically symmetric black hole solutions were derived in $(3+1)$ and $(2+1)$ dimensions \cite{Sebastiani:2010kv,Multamaki:2006zb,Hendi:2014wsa,Hendi:2014mba}, while in \cite{Multamaki:2006ym,Nashed:2020kdb,Nashed:2020tbp,Nashed:2019uyi,Nashed:2019tuk,Cembranos:2011sr} charged and rotating solutions were discussed.
  Static and spherically symmetric black hole solutions were investigated  with constant curvature,  with and without electric charge and cosmological constant in \cite{delaCruzDombriz:2009et,Hendi:2011eg,Eiroa:2020dip}. In \cite{Tang:2020sjs} curvature scalarization of black holes in $f(R)$ gravity was discussed and it was shown that the presence of a scalar field minimally coupled to gravity with a self-interacting potential can generate a rich structure of scalarized  black hole solutions, while in \cite{Tang:2019qiy} exact charged black hole solutions with dynamic curvature in D-dimensions were obtained in Maxwell-f(R) gravity theories.

 In this work we present, to the best of our knowledge,  the first exact black hole solution in $(2+1)$-dimensions of a scalar field minimally coupled to gravity in the context of $f(R)$ gravity. Without specifying the form of the $f(R)$ function, we first obtain an exact black hole solution dressed with a scalar hair with  the scalar charge to appear in the metric  and in the $f(R)$ function and discuss its thermodynamics. This solution at large distances gives a hairy BTZ black hole, and it  reduces to the BTZ black hole   when the scalar field  decouples. In a pure $f(R)$ gravity supported by the scalar field, we find an exact hairy black hole similar to the BTZ black hole with phantom hair and discuss its thermodynamics.

The work is organized as follows. In Section \ref{sect2} we derive the field equations with and without a self-interacting potential for the scalar field.
In Section \ref{sect3} we discuss hairy black hole solutions of the field equations when we have the Einstein-Hilbert term with curvature corrections and also  black hole solutions  when the $f(R)$ is purely supported by the scalar field. Finally we conclude in Section \ref{sect4}.

\section{The setup-derivation of the field equations}
\label{sect2}

We will consider  the $f(R) $ gravity theory with a scalar field minimally coupled to gravity in the presence of a self-interacting potential. Varying this action
we will look for hairy black hole solutions. We will show that if this scalar field decouples, we recover $f(R)$ gravity.
 First we will consider the case in which the scalar field does not have self-interactions.

 \subsection{Without self-interacting potential}

Consider the action
\begin{equation}
    S=\int d^3 x \sqrt{-g}\left\{\frac{1}{2\kappa}f(R) -\frac{1}{2}g^{\mu\nu}\partial_\mu \phi \partial_\nu\phi  \right\}~,\label{action0}
\end{equation}
where $\kappa$ is the Newton gravitational constant $\kappa=8\pi G$. The Einstein equations read
\begin{equation}
    f_R R_{\mu\nu}-\frac{1}{2}g_{\mu\nu}f(R)+g_{\mu\nu}\square f_R-\nabla_\mu \nabla_\nu f_R=\kappa T_{\mu\nu}~,
\end{equation}
where  $f'(R) = f_{R}$ and the energy-momentum tensor $T_{\mu\nu}$ is given by
\begin{equation}
    T_{\mu\nu}=\partial_\mu \phi \partial_\nu\phi-\frac{1}{2}g_{\mu\nu}g^{\alpha\beta} \partial_\alpha \phi \partial_\beta\phi~.
\end{equation}
The Klein-Gordon equation reads
\begin{equation}
    \square\phi=0~.
\end{equation}
We consider a spherically symmetric ansatz for the metric
\begin{equation}
    ds^2=-b(r)dt^2+\frac{1}{b(r)}dr^2+r^2d\theta^2~. \label{metr}
\end{equation}
For the metric above, the Klein-Gordon equation becomes
\begin{equation} \Box \phi = b(r)\phi''(r) + \phi'(r)\Big(b'(r) + \cfrac{b(r)}{r}\Big) =0~, \end{equation}
and takes the form of a total derivative
\begin{equation} b(r)\phi'(r)r=C~, \end{equation}
where $C$ is a constant of integration. In order to have a black hole, we require at the horizon to have  $r=r_{H} \rightarrow b(r_H)=0$. Then, $C=0$. This means that either $b(r)=0$ for any $r>0$ and no geometry can be formed, or the scalar field is constant $\phi(r) = c$. We indeed expected this behaviour, which cannot be cured with the addition of a second degree of freedom in the metric (\Ref{metr}). From the no-hair theorem \cite{Bekenstein:1971hc} we know that the scalar field should satisfy its equation of motion for the black hole geometry, thus if we multiply the Klein-Gordon equation by $\phi$ and integrate over the black hole region we have
\begin{equation} \int d^3 x  \sqrt{-g} \big(\phi \Box \phi\big) \approx \int d^3x \sqrt{-g} \nabla^{\mu}\phi \nabla_{\mu}\phi =0~ \label{no hair}, \end{equation}
where $ \approx $ means equality modulo total derivative terms. From equation (\Ref{no hair}) one can see that the scalar field is constant.

\subsection{With self-interacting potential}

We shown that if the matter does not have self-interactions then there are no hairy black holes in the $f(R)$ gravity. We then have to introduce  self-interactions for the scalar field. Consider the action
\begin{equation}
    S=\int d^3 x \sqrt{-g}\left\{\frac{1}{2\kappa}f(R) -\frac{1}{2}g^{\mu\nu}\partial_\mu \phi \partial_\nu\phi -V(\phi) \right\}~.\label{action1}
\end{equation}
The scalar field and the scalar potential obey the following conditions
\begin{equation}
    \phi\left(r\to \infty\right)=0~,\quad V\left(r\to \infty\right)=0~,\quad V\big|_{\phi=0}=0~.
\end{equation}
Varying the action (\ref{action1}) using the metric ansatz (\Ref{metr})
we get the $tt,rr,\theta\theta$ components of Einstein's equations (for $\kappa = 1$) and the Klein-Gordon equation
\begin{equation}  r \left(b'(r) f_R'(r)-f_R(r) b''(r)-f\left(r\right)+b(r) \left(2 f_R''(r)+\phi '(r)^2\right)+2 V(\phi)\right)-f_R(r) b'(r)+2 b(r) f_R'(r) =0~, \label{tt}\end{equation}
\begin{equation}  b(r) \left(r \left(-b'(r) f_R'(r)+f_R(r) b''(r)+f\left(r\right)+b(r) \phi '(r)^2-2 V(\phi)\right)+f_R(r) b'(r)-2 b(r) f_R'(r)\right)=0 ~, \label{rr} \end{equation}
\begin{equation} -r \left(2 b'(r) f_R'(r)+b(r) \left(2 f_R''(r)+\phi '(r)^2\right)+2 V(\phi)\right)+2 f_R(r) b'(r)+r f\left(r\right) =0~, \label{uu} \end{equation}
\begin{equation} \frac{\left(r b'(r)+b(r)\right) \phi '(r)}{r}+b(r) \phi ''(r)-\frac{V'(r)}{\phi '(r)}=0 ~. \label{KG}\end{equation}
The Ricci Curvature for the metric (\Ref{metr}) reads
\begin{equation} R(r) =  -\frac{2 b'(r)}{r}-b''(r)~. \label{Ricci} \end{equation}
From (\Ref{tt}) and (\Ref{rr}) equations we obtain the relation between $f_R(r)$ and $\phi(r)$
\begin{equation}  f_R''(r)+\phi '(r)^2=0~, \label{central1}\end{equation}
while the (\Ref{tt}) and (\Ref{uu}) equations yield the relation between the metric finction $b(r)$ and $f_R(r)$
\begin{equation} \left(2 b(r)-r b'(r)\right) f_R'(r)+f_R(r) \left(b'(r)-r b''(r)\right)=0 ~. \label{fb} \end{equation}
Both equations (\Ref{central1}), (\Ref{fb}) can be immediately integrated to yield
\begin{equation} f_R(r) = c_1 +c_2 r - \int \int \phi '(r)^2 dr dr ~, \label{CENTRAL} \end{equation}
\begin{equation} b(r) = c_3 r^2 - r^2 \int \cfrac{K}{r^3 f_R(r)}dr ~ \label{fb1}\end{equation}
where $c_1, c_2, c_3$ and $K$ are constants of integration. We can also integrate the Klein-Gordon equation
\begin{equation} V(r) = V_0 + \int \frac{r b'(r) \phi '(r)^2+rb(r) \phi '(r) \phi ''(r)+b(r) \phi '(r)^2}{r}dr ~. \label{Klein} \end{equation}
Equation (\Ref{CENTRAL}) is the central equation of this work. First of all, we recover General Relativity for the vanishing of scalar field and for $c_1=1, c_2=0$. We stress the fact that in $f(R)$ gravity we are able to derive non-trivial configurations for the scalar field with  one degree of freedom as can be seen in the metric  (\Ref{metr}). This is not the case in the context of General Relativity, as it is discussed in \cite{Chan:1996rd}. There we can see that a second degree of freedom (equation (4) in \cite{Chan:1996rd}) must be added for the existence of  non-trivial solutions for the scalar field. Here, the fact of non-linear gravity makes $f_R \neq const.$, and therefore we can have a one degree of freedom metric.  The integration constants $c_1$ and $c_2$ have physical meaning. $c_1$ is related with the Einstein-Hilbert term, while $c_2$ is related to possible (if $c_2 \neq 0 $) geometric corrections to General Relativity that are encoded in $f(R)$ gravity. The last term of this equation is related directly to the scalar field. This means that the matter not only modifies the curvature scalar $R$ but also the gravitational model $f(R)$.

\section{Black hole solutions}
\label{sect3}
In this section we will discuss the cases where $c_1=1, c_2=0$ and $c_1=c_2=0$ for a given scalar field configuration. For the second case
to satisfy observational and thermodynamical constraints we  will introduce a phantom scalar field and we will reconstruct the $f(R)$ theory, looking for
black hole solutions.

\subsection{$c_1=1, c_2=0$}

Equations (\Ref{CENTRAL}), (\Ref{fb1}) and (\Ref{Klein}) are three independent equations for the four unknown functions of our system, $f_R,\phi,V,b$, hence we have the freedom to fix one of them and solve for the others. We fix the scalar field configuration as
\begin{equation} \phi (r) = \sqrt{\cfrac{A}{r+B}} \label{scalar} ~, \end{equation}
where $A$ and $B$ are some constants with unit $[L]$, the scalar charges. We now obtain from equation (\Ref{CENTRAL}) $f_R(r)$
\begin{equation} f_R(r) = 1-\frac{A}{8 (B+r)}  ~, \end{equation}
where we have set $c_2=0$ and $c_1=1$. Therefore, we expect that, at least in principle, a pure Einstein-Hilbert term will be generated if we integrate $f_R$ with respect to the Ricci scalar. \\
Now, from equation (\Ref{fb1}) we obtain the metric function
\begin{equation} b(r)=c_3 r^2-\frac{4 B K}{A-8 B}-\frac{8 A K r}{(A-8 B)^2}-\cfrac{64 A K r^2 }{(A-8 B)^3}\ln \left(\cfrac{8 (B+r)-A}{r}\right)~. \label{b} \end{equation}
The metric function is always continuous for positive $r$ when the scalar charges satisfy $0<A<8B$. Here we show its asymptotic behaviors at the origin and space infinity
\begin{eqnarray}
b(r \rightarrow 0) &=& -\frac{4 B K}{A-8 B}-\frac{8 A K r}{(A-8 B)^2} +c_3 r^2  +\frac{64 A K r^2}{(A-8 B)^3} \ln \left(-\frac{r}{A-8 B}\right) + \mathcal{O}(r^{3})~,\\
b(r \rightarrow \infty)&=&  \frac{K}{2}+\frac{A K}{24 r} -r^2 \Lambda_{\text{eff}} +\mathcal{O}(r^{-2})~,
\end{eqnarray}
where the effective cosmological constant of this solution is generated from the equations can be read off
\begin{equation} \Lambda_{\text{eff}} =- c_3+\frac{192 A K \ln (2)}{(A-8 B)^3}~. \end{equation}

 It is important to discuss the asympotic behaviours of the metric function. At large distances, we can see that we obtain the BTZ black hole where the scalar charges appear in the effective cosmological constant of the solution. Corrections in the structure of the metric appear as $\mathcal{O}(r^{-n})$ (where $n\ge1$) terms and are completely supported by the scalar field. At small distances we can see that the metric function has a completely different behaviour from the BTZ black hole. Besides the constant and $\mathcal{O}(r^2)$ terms there are present $\mathcal{O}(r)$ and $\mathcal{O}(r^2\ln(r))$ terms that have an impact on the metric for small $r$. Our findings are in agreement with the work \cite{Tang:2020sjs} where in four dimensions Schwarzchild black holes are obtained at infinity with a scalarized mass term while at small distances a rich structure of black holes is unveiled.   This is expected since at small distances the Ricci curvature becomes strong and therefore changing the form of spacetime.
The Ricci scalar and the Kretschmann scalar are both divergent at the origin
\begin{eqnarray}
R\left(r\to 0\right)&=&\frac{16 A K}{r (A-8 B)^2}+\mathcal{O}\left(\ln{r}\right)~,\\
K\left(r\to 0\right)&=&\frac{128 K^2 A^2}{r^2 (A-8 B)^4}+\mathcal{O}\left(\frac{1}{r}\ln{r}\right)~,
\end{eqnarray}
indicating a singularity at $r=0$. As a consistency check for $A=0$ we indeed obtain the BTZ \cite{BTZ1} black hole solution
\begin{equation} b(r) = c_3 r^2 + \frac{K}{2}~,\end{equation}
which means that for vanishing scalar field we go back to General Relativity. Hence the solution (\Ref{b}) can be regarded as a scalarized version of the BTZ black hole in the context of $f(R)$ gravity.

Now we solve the expression of the potential from the Klein-Gordon equation
\begin{multline}
V(r)= \frac{1}{8 A B^2 (A-8 B)^3 (B+r)^3}\\
\bigg(B (4 A^4 (-B^2 (K-18 c_3 r^2)+36 B^3 c_3 r+12 B^4 c_3-4 B K r-2 K r^2)-64 A^3 B (r^2 (9 B^2 c_3+K) \\
+B r (18 B^2 c_3+K)+6 B^4 c_3 )+256 A^2 B (B (6 r^2 (B^2 c_3+K)+2 B r (6 B^2 c_3+5 K)+4 B^4 c_3+3 B^2 K)+\\
30 K \ln (2) (B+r)^3)-A^5 B c_3 (2 B^2+6 B r+3 r^2)+64 B K (-A^3 (2 B^2+6 B r+3 r^2) \ln (\frac{r}{8 (B+r)-A})\\
-8 (5 A^2-32 A B+64 B^2) (B+r)^3 \ln (8 (B+r)-A))-4096 A B^2 K (B+r)^2 (12 \ln (2) (B+r)+B) \\
+98304 B^3 K \ln (2) (B+r)^3)-8 A^2 K (A^2-32 A B+64 B^2) (B+r)^3 \ln (r)+8 K (A-8 B)^4 (B+r)^3 \ln (B+r)\bigg)~, \end{multline}
the asymptotic behaviors of which are
\begin{eqnarray}
V(r \rightarrow 0)&=& -\frac{K \ln (r)}{B^2 (A-8 B)} + \mathcal{O}(r^0)~, \\
V(r \rightarrow \infty)&=& \frac{3 A \left(24 A^2 B c_3-A^3 c_3-192 A \left(B^2 c_3-K \ln (2)\right)+512 B^3 c_3\right)}{8 r (A-8 B)^3} +\mathcal{O}\left(\frac{1}{r^2}\right)~.
\end{eqnarray}
To ensure  that the potential vanishes at  space infinity, we need to set the integration constant  $V_0$ at (\Ref{Klein}) equal to
\begin{equation}
    V_0=\frac{192 K\ln{2} \left(5 A^2-32 A B+64 B^2\right)}{A (A-8 B)^3}~.
\end{equation}
In addition, there is a mass term in the potential that has the same sign with the effective cosmological constant
\begin{equation} m^2 = V''(\phi=0) = \frac{3}{4} \left(\frac{192 A K \ln (2)}{(A-8 B)^3}-c_3\right) = \frac{3}{4}\Lambda_{\text{eff}}~, \end{equation}
which satisfies the Breitenlohner-Freedman bound in three dimensions \cite{Breitenlohner:1982jf, Mezincescu:1984ev}, ensuring the stability of AdS spacetime under perturbations  if we are working in the  AdS spacetime.

Substituting the obtained configurations into one of the Einstein equations we can solve for $f(r)$
\begin{multline}
f(r) = \frac{1}{A B^2 r (A-8 B)^3 (A-8 (B+r))}\bigg[ B (192 B K r \ln (2) (5 A^2-32 A B+64 B^2) (A-8 (B+r))+A (A-8 B)^2 (16 B c_3 r^2 (A-8 B) \\ -2 B c_3 r (A-8 B)^2+8 K r (A+8 B)-A K (A-8 B)))+A^2 K r (-(A^2-32 A B+64 B^2)) \ln (r) (A-8 (B+r)) \\+K r (8 (B+r)-A) \left(64 B^2 ((5 A^2-32 A B+64 B^2) \ln (8 (B+r)-A)+2 A^2 \ln (\frac{r}{8 (B+r)-A}))-(A-8 B)^4 \ln (B+r)\right)\bigg]. \label{f(r)1}
\end{multline}
On the other side, the Ricci scalar can be calculated from the metric function
\begin{equation} R(r)=\frac{16 A K \left(-36 r (A-8 B)+(A-8 B)^2+192 r^2\right)}{r (A-8 B)^2 (A-8 (B+r))^2}+\frac{384 A K }{(A-8 B)^3}\ln \left(\cfrac{8 (B+r)-A}{r}\right)-6 c_3.~ \label{ricci1}\end{equation}
As one can see it is difficult to invert the Ricci scalar and solve the exact form of $f(R)$, though we have the expressions of $R(r)$, $f(r)$ and $f_R(r)$. Nevertheless we can still obtain the asymptotic $f(R)$ forms by studying their asymptotic behaviors
\begin{eqnarray}
f\left(r \to \infty\right) &=& -\frac{A K (A-8 B)}{128 r^4}+\frac{768 A K \ln (2)}{(A-8 B)^3}-4 c_3 +\mathcal{O}\left(\frac{1}{r^5}\right)~,\\
R\left(r \to \infty\right) &=&-\frac{A K (A-8 B)}{128 r^4}+\frac{1152 A K \ln (2)}{(A-8 B)^3}-6 c_3 +\mathcal{O}\left(\frac{1}{r^5}\right)~,\\
f\left(r \to 0 \right) &=& -\frac{2 A K}{(A-8 B)Br} +\mathcal{O}\left(\ln{r}\right)~,\\
R\left(r \to 0 \right) &=&\frac{16 A K}{r (A-8 B)^2}+\mathcal{O}\left(\ln{r}\right)~,
\end{eqnarray}
which leads to
\begin{eqnarray}
f(R) &\simeq& R +2c_3-\frac{384 A K \ln (2)}{(A-8 B)^3} = R -2\Lambda_{\text{eff}}~, \quad\quad r\to \infty~, \label{f(R)I}\\
f(R) &\simeq& R \left(1-\frac{A}{8 B}\right)~, \quad\quad r\to 0 ~.\label{f(R)O}
\end{eqnarray}

\begin{figure}[H]
\centering
 \includegraphics[width=.40\textwidth]{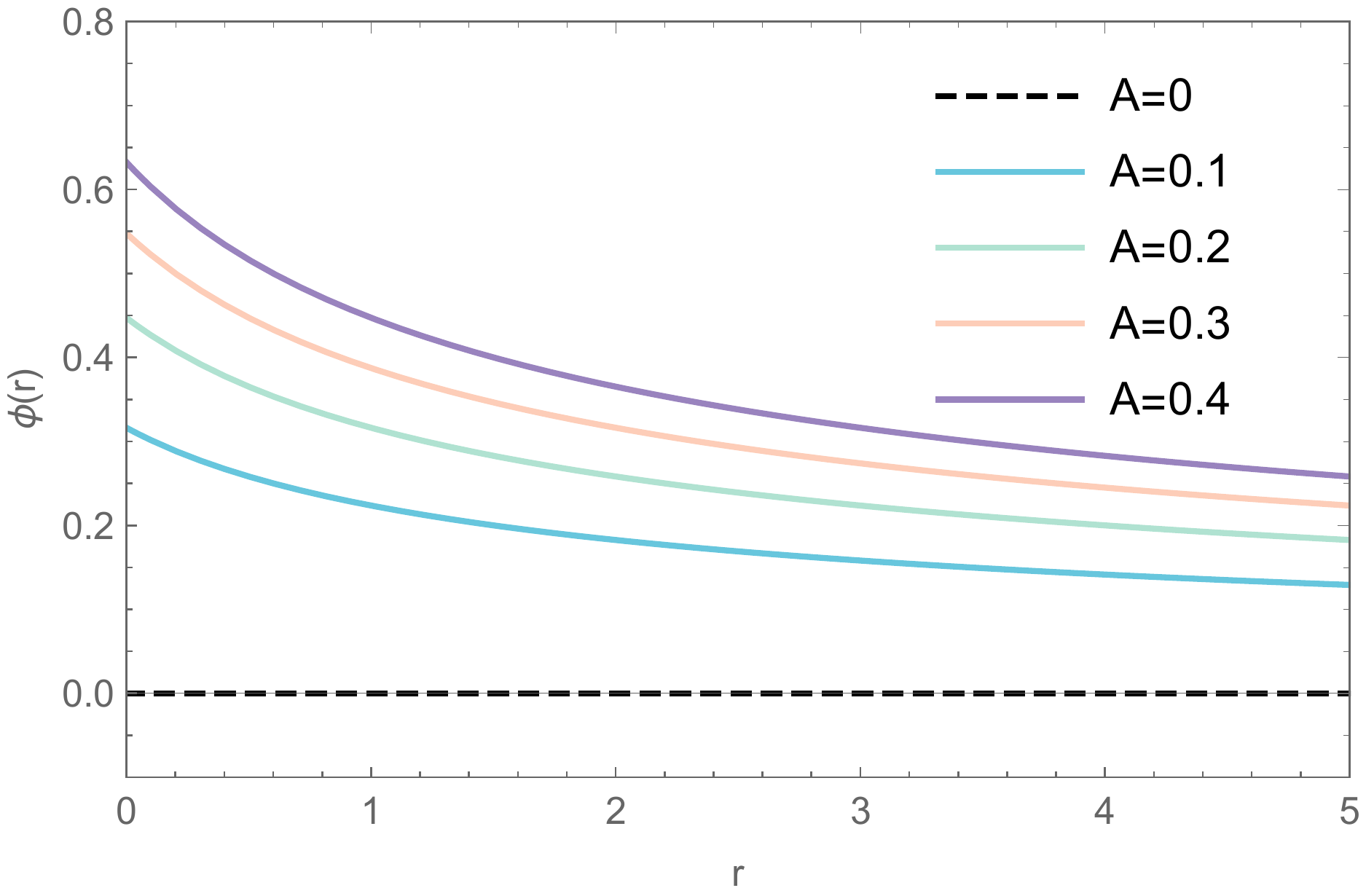}
 \includegraphics[width=.40\textwidth]{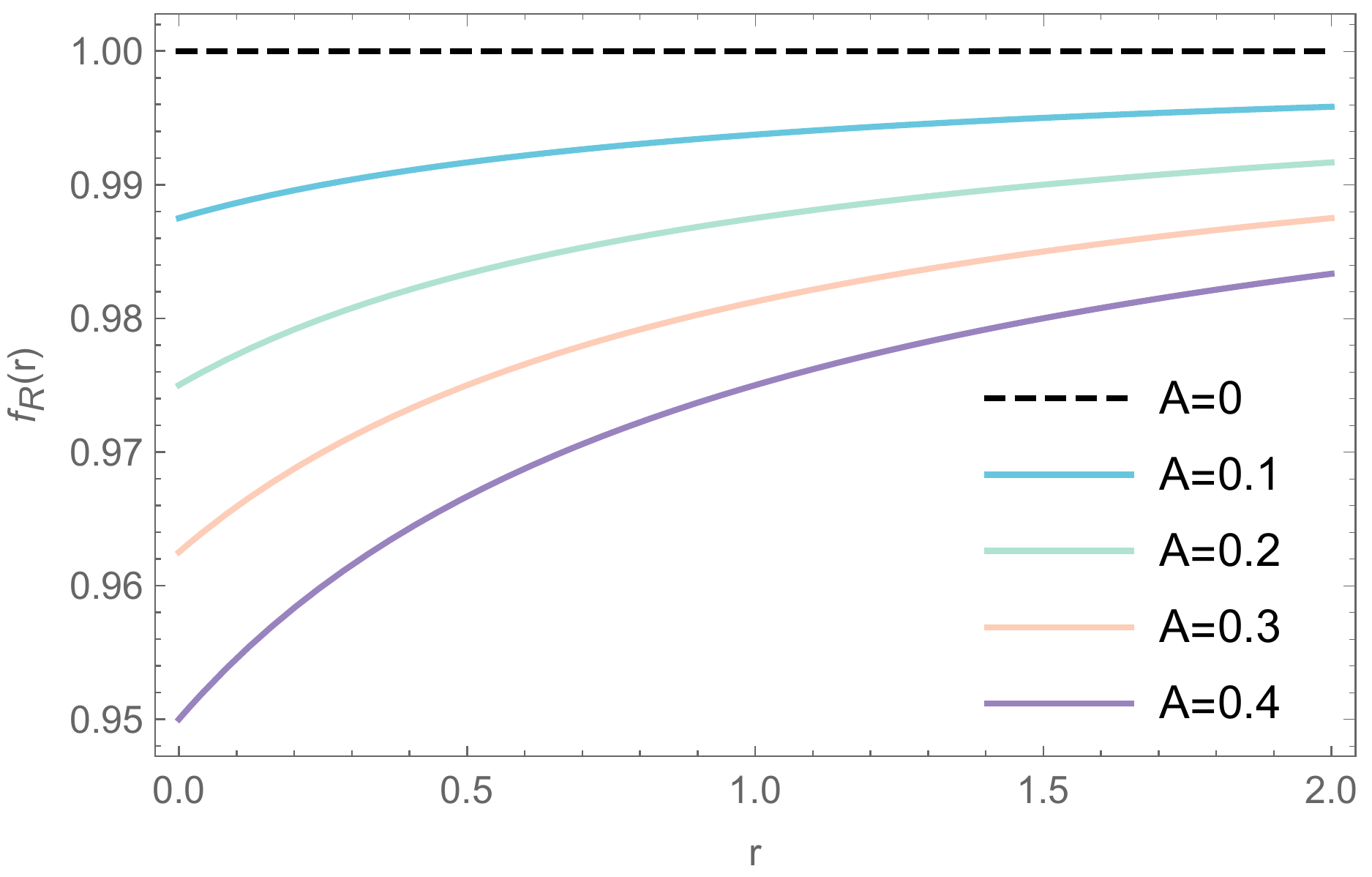}
 \includegraphics[width=.40\textwidth]{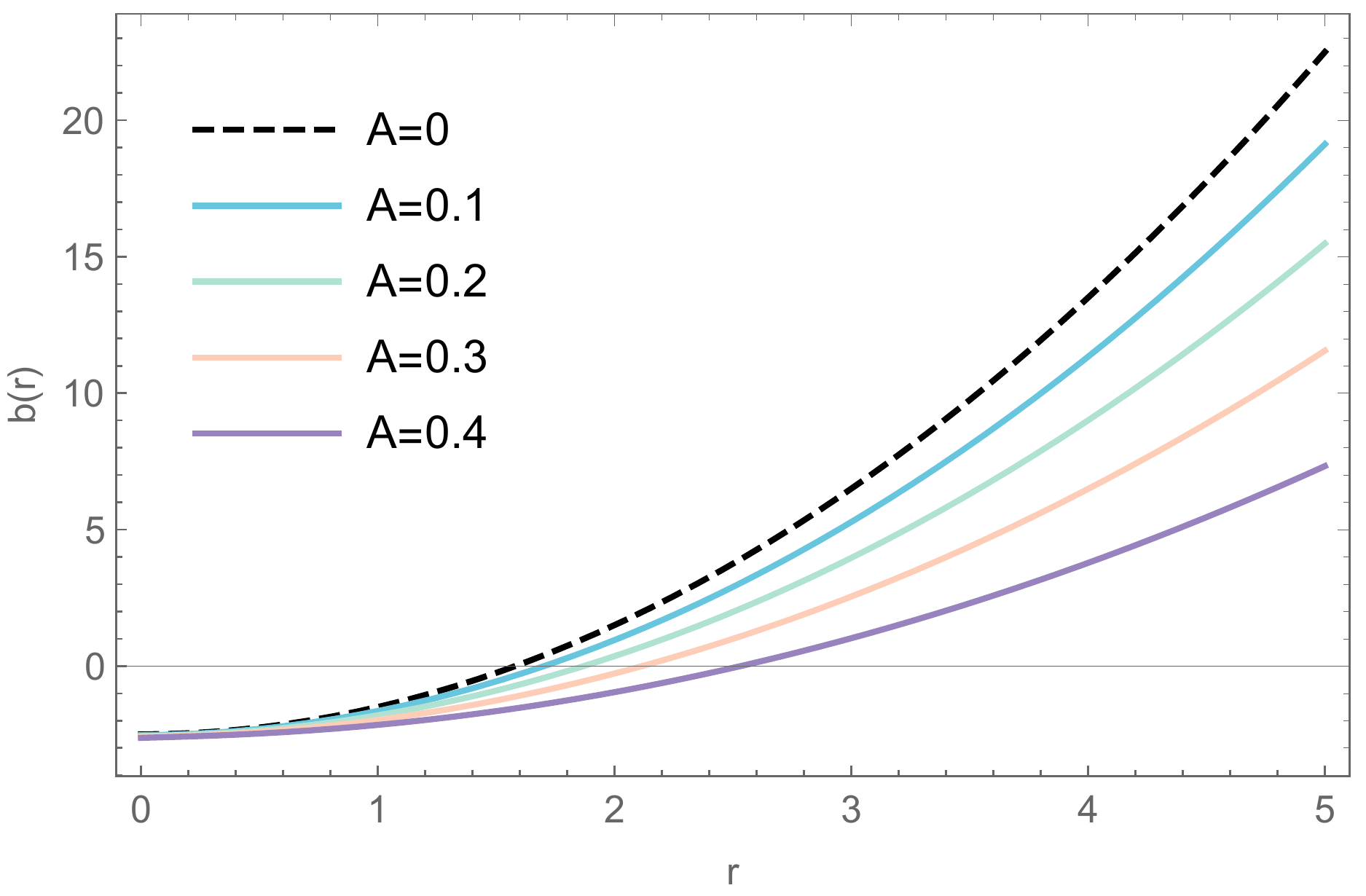}
 \includegraphics[width=.40\textwidth]{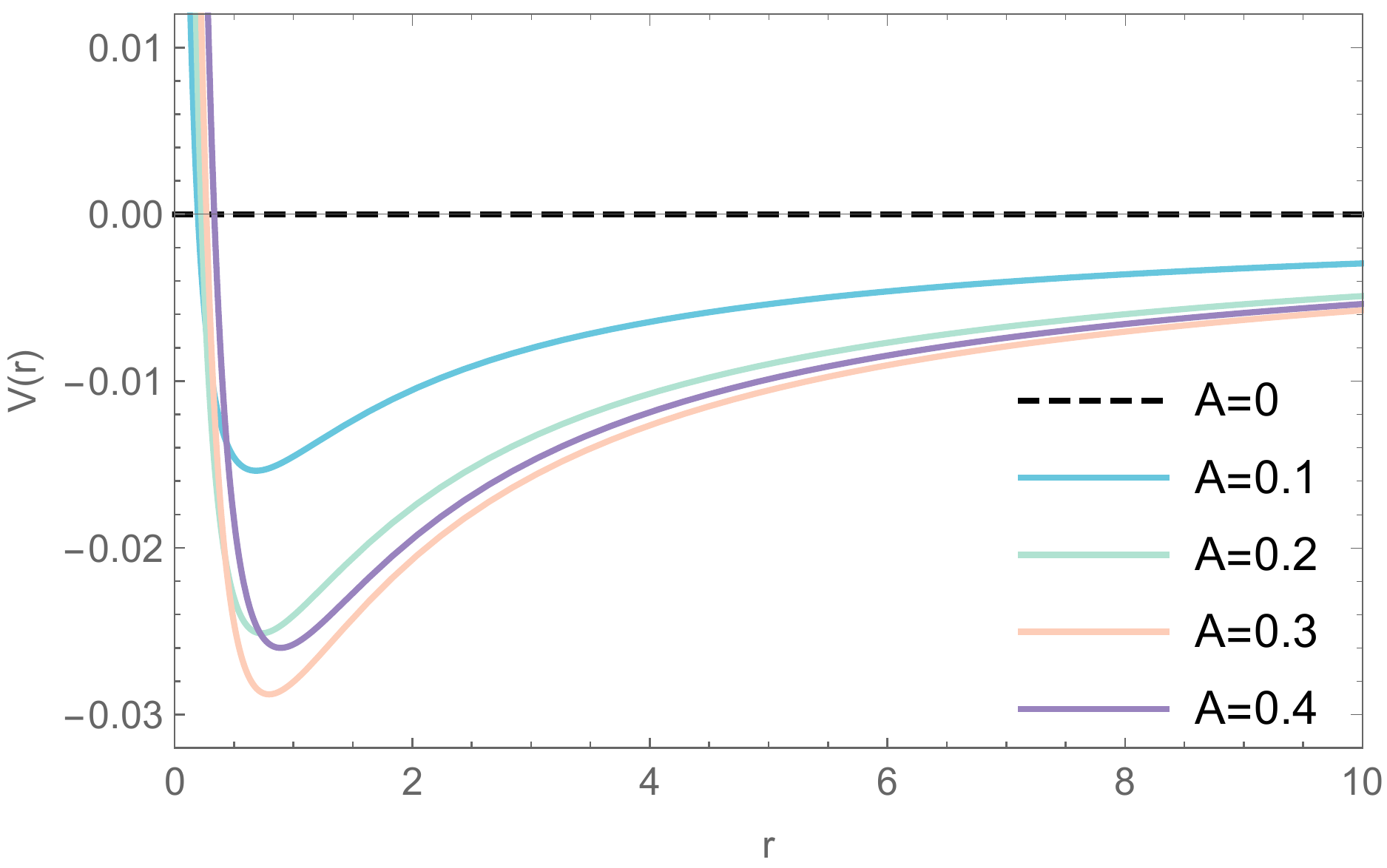}
\includegraphics[width=.40\textwidth]{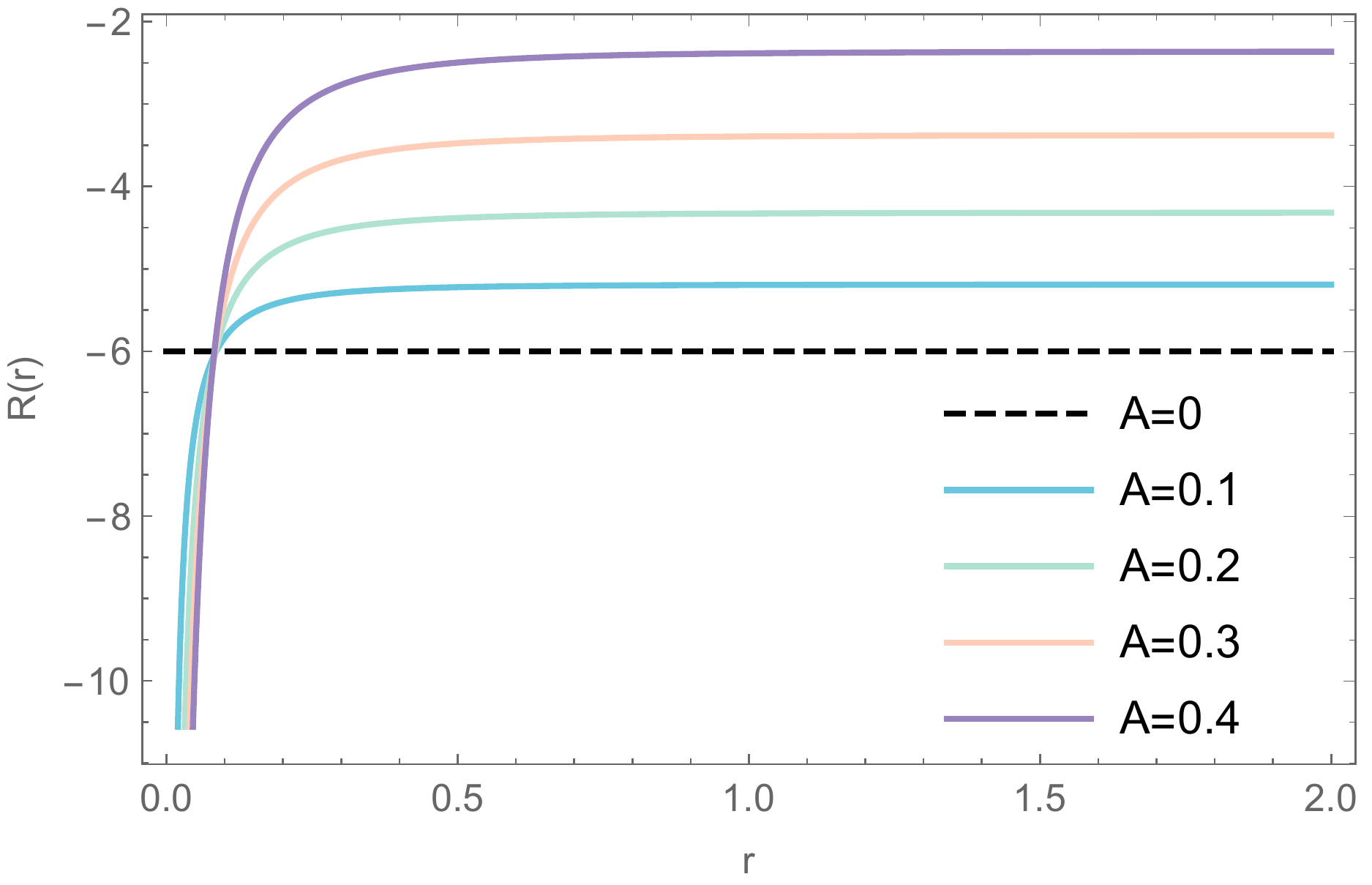}
 \includegraphics[width=.40\textwidth]{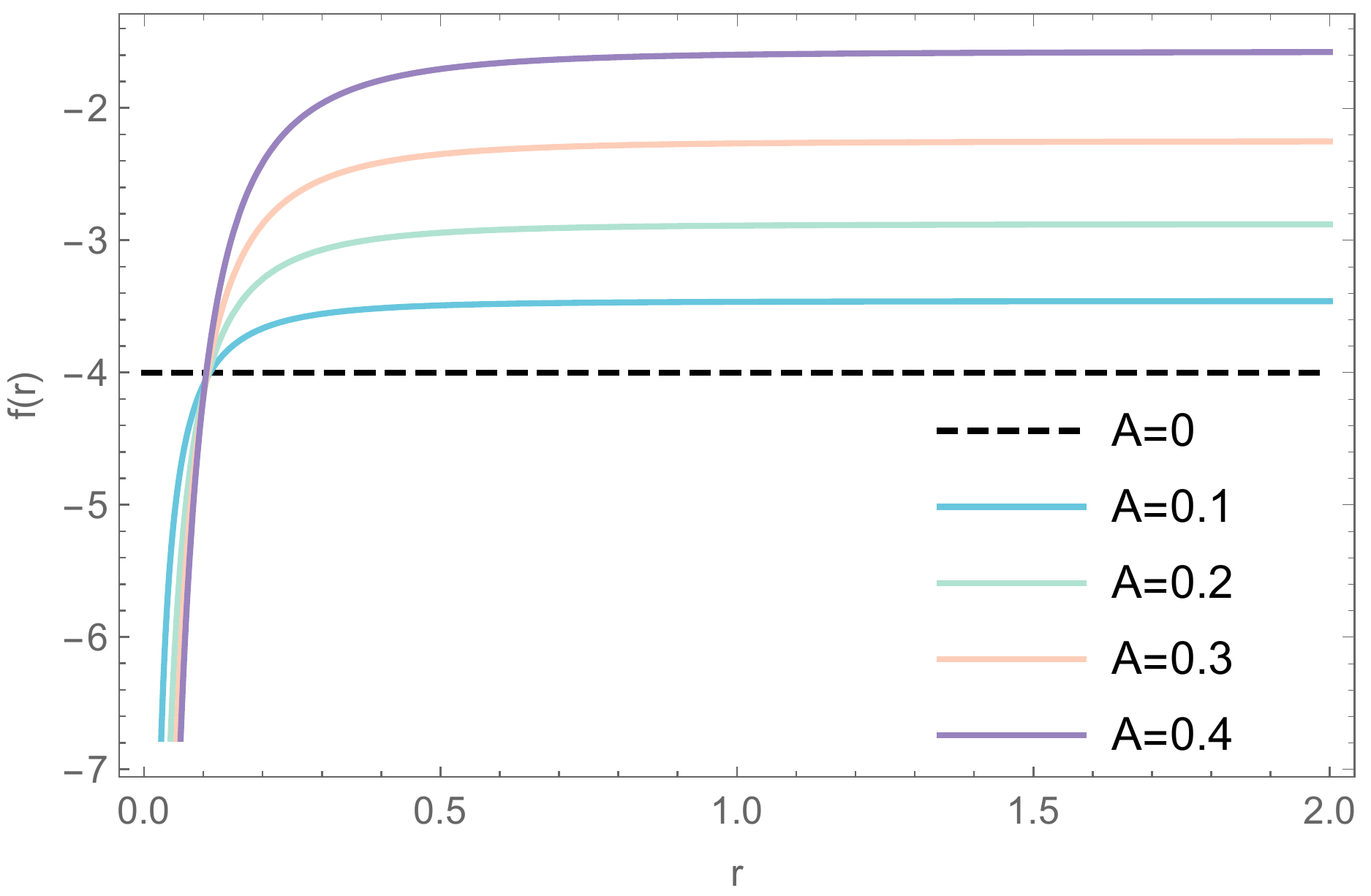}
\caption{All the physical quantities of the AdS black holes are plotted with different scalar charges $A$, where other parameters have been fixed as $B=1$, $K=-5$ and $c_3=1$.} \label{fig:AdSBH1}
\end{figure}

\begin{figure}[H]
\centering
 \includegraphics[width=.40\textwidth]{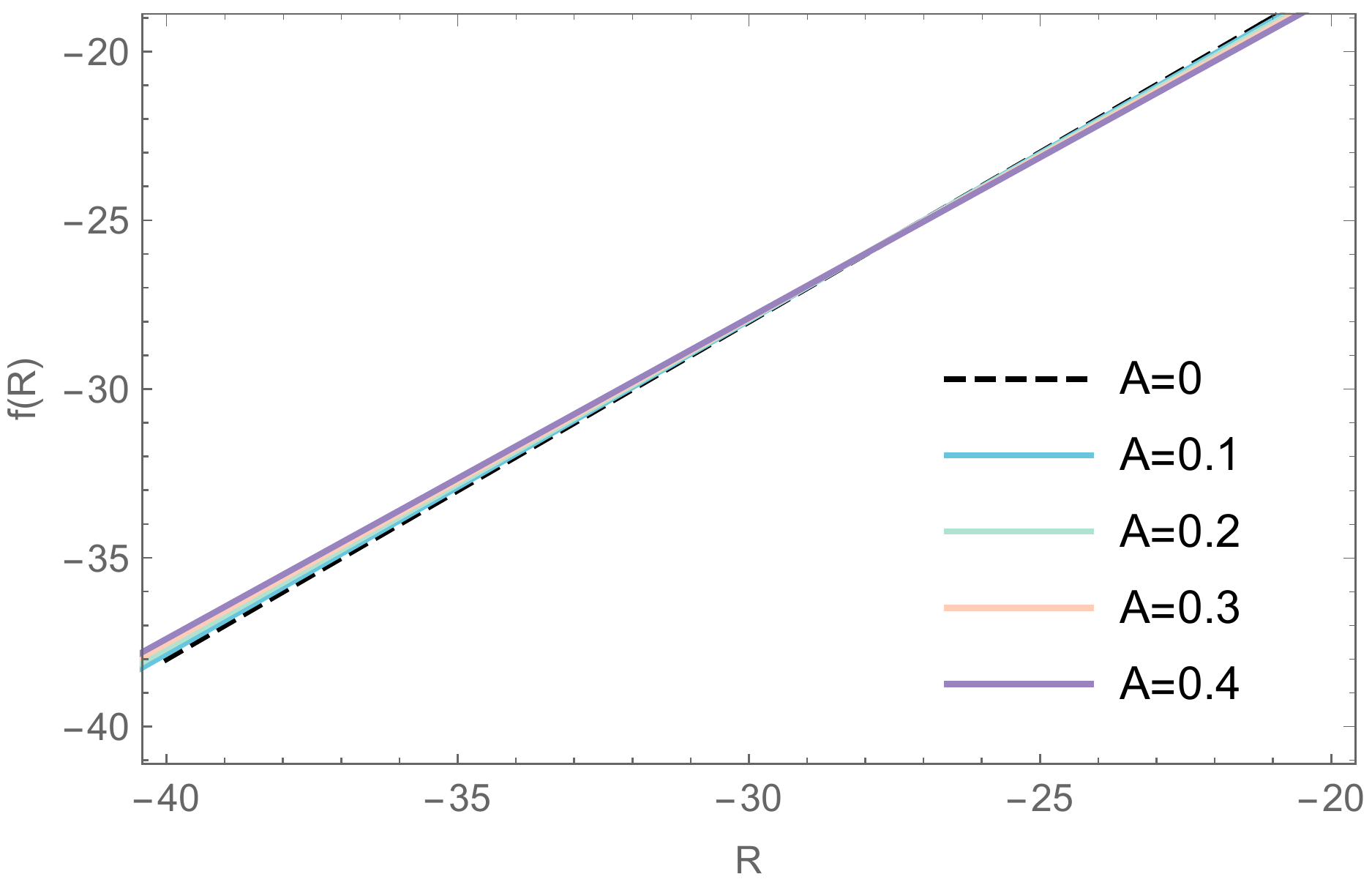}
 \includegraphics[width=.40\textwidth]{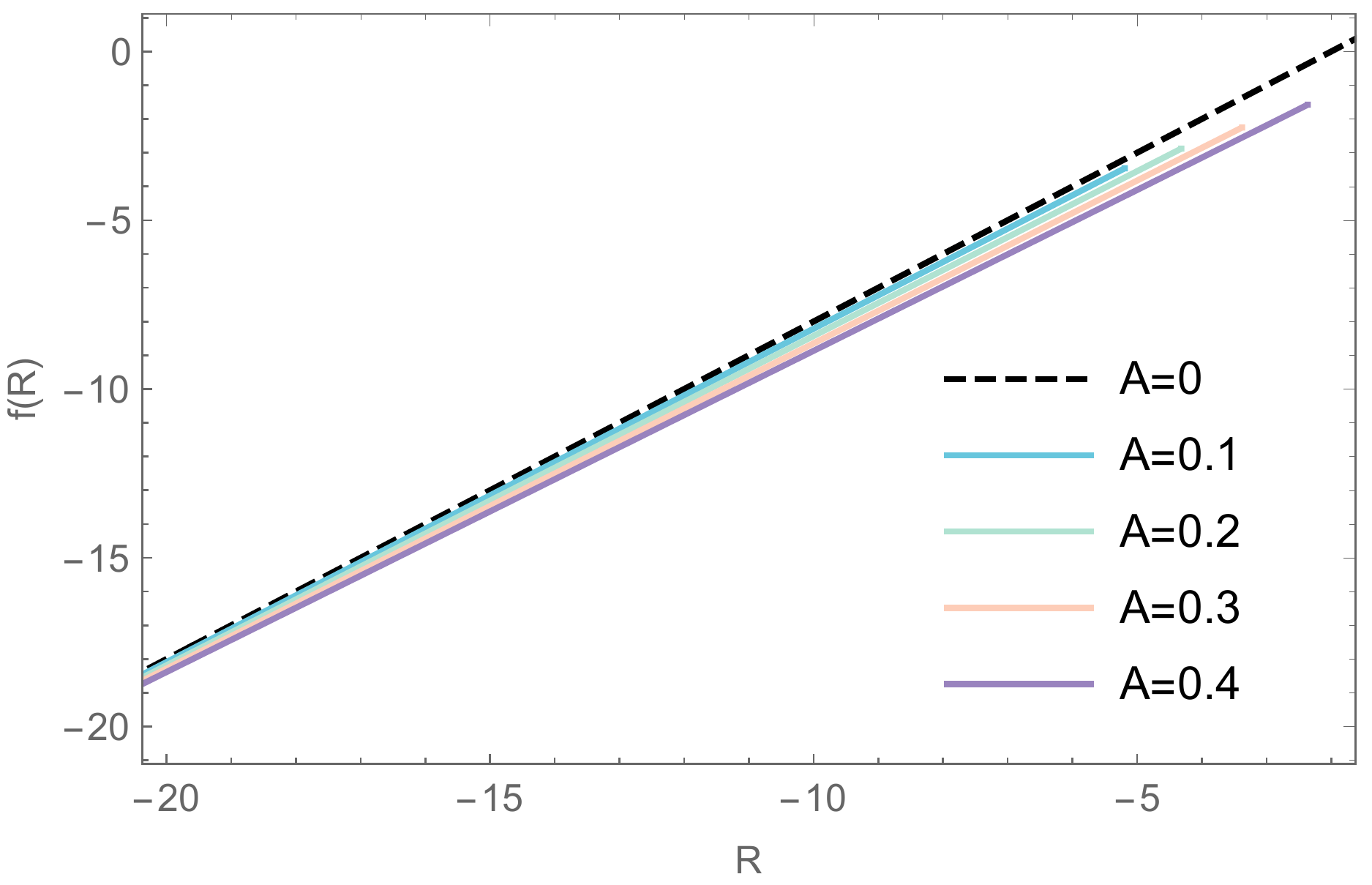}
 \caption{The $f(R)$ function. The black dashed line represents the Einstein Gravity $f(R)=R-2\Lambda_{\text{eff}}$, where other parameters have been fixed as $B=1$, $K=-5$ and $c_3=1$.} \label{fig:AdSBH2}
\end{figure}

The fact that the Ricci scalar contains logarithmic terms prevents us from obtaining the non-linear corrections near the origin, where we expect the modified part of the $f(R)$ model to be stronger, since it is supported by the existence of the scalar field and the scalar field takes its maximum value for $r=0 \to \phi(0) = \sqrt{A/B}$. To avoid the tachyonic instability, we check the Dolgov-Kawasaki stability ctiterion \cite{Hendi:2014mba} which states that the second derivative of the gravitational model $f_{RR}$ must be always positive \cite{DeFelice:2010aj,Bertolami:2009cd,Faraoni:2006sy}. Using the chain rule
\begin{equation} f_{RR} = \cfrac{df_R(R)}{dR} = \cfrac{df_R(r)}{dr} \cfrac{dr}{dR} = \cfrac{f'_R(r)}{R'(r)} = -\frac{r^2 (A-8 (B+r))^3}{128 K (A-8 B) (B+r)^2}~, \end{equation}
we can see that the above expression is always positive for $K<0$ when the continuity condition $0<A<8B$ is considered. So far we have not imposed any condition on $c_3$, therefore the spacetime might be asympotically AdS or dS depending on the value of parameter $c_3$
\begin{eqnarray}
c_3&>&\frac{192 A K \ln (2)}{(A-8 B)^3}>0 \hspace{0.5cm} \text{asympotically~AdS~}, \\
c_3&<&\frac{192 A K \ln (2)}{(A-8 B)^3} \hspace{0.5cm} \text{asympotically~dS~}.
\end{eqnarray}
We can prove that the metric function has at most one root, which can not describe a dS black hole. For the asympotically AdS spacetime, the condition $K<0$ gives an AdS black hole solution while the condition $K>0$ gives the pure AdS spacetime with a naked singularity at origin. For the asympotically dS spacetime, the condition $K>0$ gives a pure dS spacetime with a cosmological horizon. Therefore pure AdS or dS spacetime described by this solution suffers from the tachyonic instability, only AdS black holes can survive from this instability.  We plot all the physical quantities of the AdS black holes in FIG. \ref{fig:AdSBH1} and FIG. \ref{fig:AdSBH2}.  In FIG. \ref{fig:AdSBH1} we plot the metric function, the potential, the scalar field, the Ricci scalar, the $f(r)$ and  $f_R$ functions along with the $A=0$ (BTZ black hole) case in order to compare them. In FIG. \ref{fig:AdSBH2} we plot the $f(R)$ model along with $f(R) = R-2 \Lambda_{\text{eff}}$ in order to compare our model with Einstein's Gravity. For FIG. \ref{fig:AdSBH2} we used the expression for the Ricci scalar (\Ref{ricci1}) for the horizontal axes and the expression for $f(r)$ (\ref{f(r)1}) for the vertical axes.

From FIG. \ref{fig:AdSBH1} and FIG. \ref{fig:AdSBH2} we can see that the existence of scalar charge $A$ makes the solution deviate from the GR solution, and the stronger the scalar charge is, the larger it deviates. The figure of the metric function shows that the hairy solution with stronger scalar charge has larger radius of the event horizon, while its influence on the curvature is qualitative, from constant to dynamic, with a divergence appearing at origin. The scalar charge also modifies the $f(R)$ model and the potential to support such hairy structures, where the potential develops a well near the origin to trap the scalar field providing the right matter concentration for a hairy black hole to be formed. For the $f(R)$ model, the scalar charge only sets aside a small distance with the Einstein Gravity while the slope changes little, indicating our $f(R)$ model is very close to Einstein Gravity. We can see that even slight deviations from General Relativity can support hairy structures. The asymptotic expressions (\ref{f(R)I}) (\ref{f(R)O}) tell us that at large scale the scalar field only modifies the effective cosmological constant while at small scale the slope of $f(R)$ can also be modified, which agrees with the figure of $f(R)$.

\begin{figure}[H]
\centering
	\includegraphics[width=.40\textwidth]{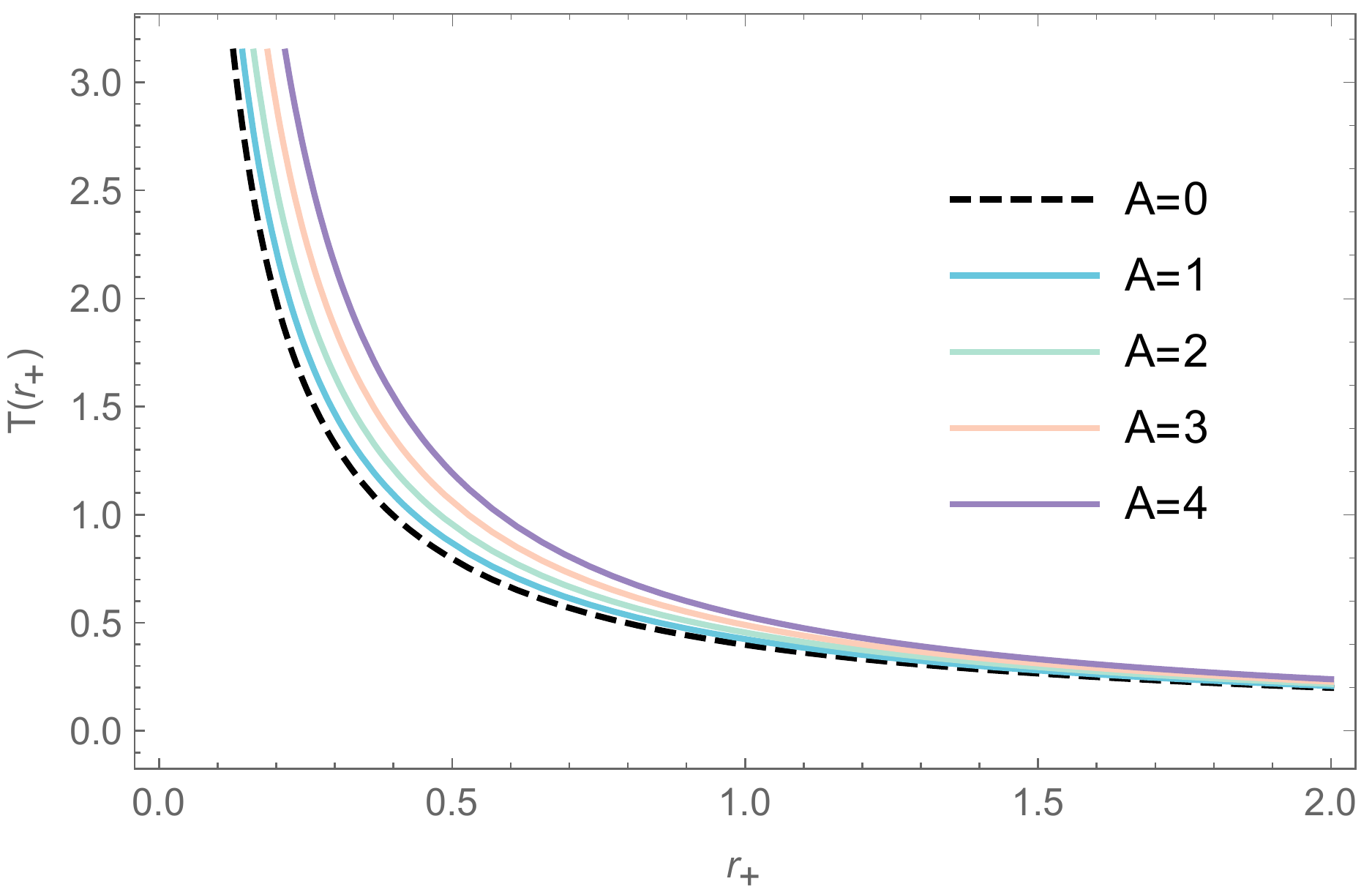}
	\includegraphics[width=.40\textwidth]{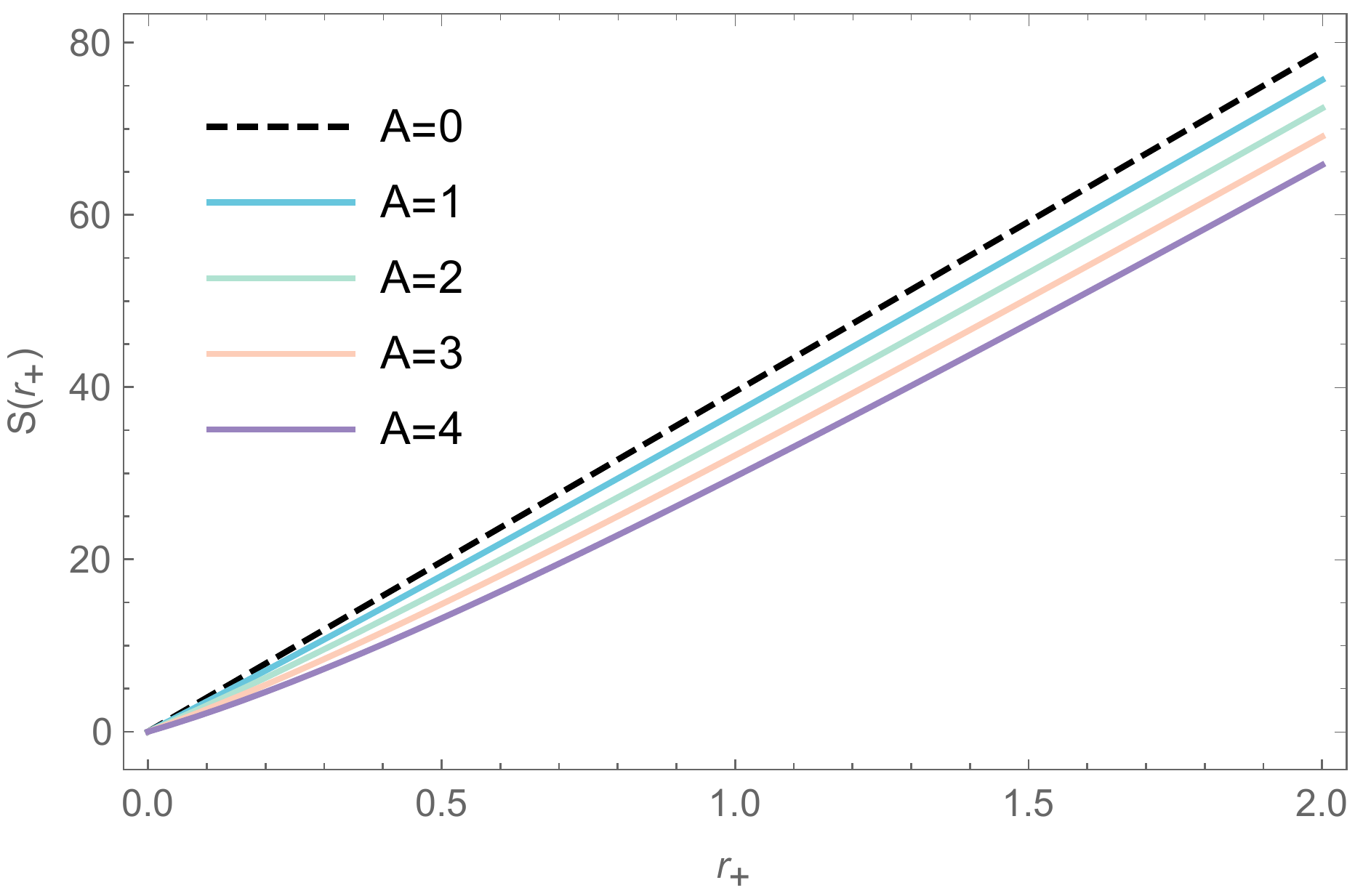}
\caption{The Hawking temperature and Bekenstein-Hawking entropy are plotted with different scalar charges $A$, where other parameters have been fixed as $B=1$ and $K=-5$.
}. \label{TS1}
\end{figure}

\begin{figure}[H]
\centering
	\includegraphics[width=.40\textwidth]{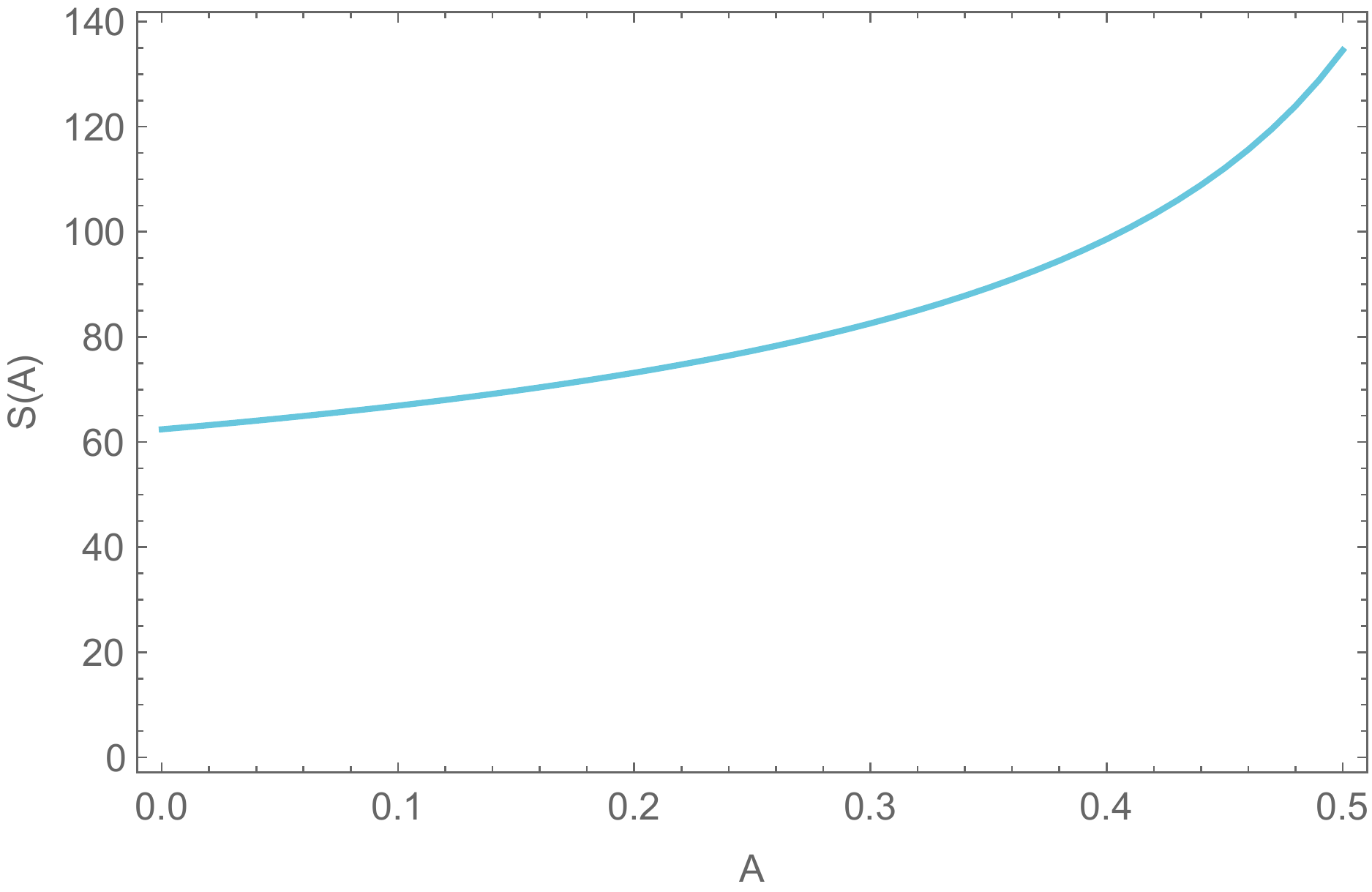}
\caption{The Bekenstein-Hawking entropy as a function of the scalar charge $A$, where other parameters have been fixed as $B=1$, $K=-5$ and $c_3=1$.
}. \label{TS2}
\end{figure}

Next we study the thermodynamics of this solution. The Hawking temperature and Bekenstein-Hawking entropy are defined as \cite{Akbar:2006mq, Camci:2020yre}
\begin{eqnarray}
T(r_+) &=& \cfrac{b'(r_+)}{4\pi}=\frac{2 K (B+r_+)}{\pi  r_+ (A-8 (B+r_+))}, \label{temperature} \\
S(r_+) &=& \frac{\mathcal{A}}{4G}f_{R}(r_+) =4\pi^2 r_+f_{R}(r_+)= 4\pi^2 r_+ \left(1-\frac{A}{8 (B+r_+)}\right), \label{entropy}
\end{eqnarray}
where $r_+$ is the radius of the event horizon of the AdS black hole and $A=2\pi r_+$ is the area of the event horizon, where the gravitational constant $G$ equals $1/8\pi$ since we've set $8\pi G=1$. Here in the first expression we have already used $r_+$ to replace the parameter $c_3$. It is clear that the Hawking temperature and Bekenstein-Hawking entropy are both positive for $K<0$ and $0<A<8B$. We present their figures in FIG. \ref{TS1}.
FIG. \ref{TS1} shows that for the same radius of the event horizon, the hairy black hole solution owns higher Hawking temperature but lower Bekenstein-Hawking entropy. However, with fixed parameters $B, c_3$ and $K$, the hairy black hole solution has larger radius of the event horizon, therefore, we plot the entropy inside the event horizon as a function of the scalar charge $A$ in FIG. \ref{TS2} to confirm if the hairy solution is thermodynamically preferred or not. The fact is that hairy black hole solution is thermodynamically preferred, which owns higher entropy than its corresponding GR solution, BTZ black hole, and the entropy grows with the increase of the scalar charge $A$. It can be easily understood that the participation of the scalar field gains more entropy for the black hole.

\subsection{Exact Black Hole Solution with Phantom Hair}

In the previous section, we have set $c_1=1$ and $c_2=0$, therefore the $f(R)$ model  consists of the pure Einstein-Hilbert term and corrections that arise from the existence of the scalar field. We have shown that with the vanishing of scalar field, we obtain the well known results of General Relativity, the  BTZ black hole \cite{BTZ1}.

We will now discuss the possibillity that the scalar field, purely supports the $f(R)$ model by setting $c_1=c_2 =0$. From equation (\ref{CENTRAL}) we can see that due to the $\mathcal{O}(r^{-n})$ (where $n>0$) nature of the scalar field and the double integration, there will be regions where $f_R <0$. For example for our scalar profile (\Ref{scalar}) the $f_R$ turns out to be
\begin{equation} f_R(r) = -\frac{A}{8 (B+r)}~, \end{equation}
which is always negative for $A, B>0$. With this form of $f_R$ one can derive an exact hairy black hole solution similar to a hairy BTZ black hole which however has negative entropy as can be seen from the relation   (\ref{entropy}).

It is clear that a sign reversal of $f(R)$ can fix the negative entropy problem. As a result, the sign reversal of other terms in the action is also required, which leads to a phantom scalar field instead of the regular one. This comes in agrement with recent observational results which they require that at the early universe to explain the equation of state $w<-1$ phantom energy is needed to support the cosmological evolution    \cite{Caldwell:1999ew, Zhang:2017tbf, Bronnikov:2005gm}. As it will be shown in the following, in the pure $f(R)$ gravity theory the curvature acquires non-linear correction terms which makes the curvature stronger as it is expected in the early universe.

Hence, we consider the following action
\begin{equation}
    S=\int d^3 x \sqrt{-g}\left\{\frac{1}{2\kappa} f(R) + \frac{1}{2} g^{\mu\nu}\partial_\mu \phi \partial_\nu\phi -V(\phi) \right\}~, \label{action2}
\end{equation}
which is the action (\Ref{action1}) but the kinetic energy of the scalar field comes with the positive sign which corresponds to a phantom scalar field instead of the regular one. Under the same metric ansatz (\Ref{metr}), equation (\Ref{central1}) now becomes
\begin{equation}  f_R''(r)-\phi '(r)^2=0~,\end{equation}
and by integration
\begin{equation} f_R(r) =\int \int \phi '(r)^2 dr dr ~,\end{equation}
having set $c_1=0$ and $c_2=0$. With the same profile of the scalar field, the solution of this action becomes
\begin{eqnarray}
\phi (r)&=&\sqrt{\frac{A}{B+r}}~,\\
f_R(r)&=&\frac{A}{8 (B+r)}~,\\
b(r) &=& \frac{4 B K}{A}+\frac{8 K r}{A}-\Lambda  r^2~, \label{b_ph}\\
R(r) &=& 6 \Lambda -\frac{16 K}{A r}~,\\
V(r) &=& \frac{B (A B \Lambda +4 K)}{8 (B+r)^3}-\frac{3 A B \Lambda +8 K}{8 B (B+r)}-\frac{K}{B^2} \ln{\left(\frac{B+r}{r}\right)}~,\\
f(r) &=& -\frac{2 K}{B r}+\frac{2 K }{B^2}\ln{\left(\frac{B+r}{r}\right)}~,\\
f(R) &=& \frac{A R}{8 B}-\frac{3 A \Lambda }{4 B}+\frac{2 K}{B^2} \ln \left(\frac{6 A B \Lambda -A B R+16 K}{16 K}\right)~,\\
V(\phi) &=& -\frac{K \phi ^2}{A B}-\frac{3 \Lambda  \phi ^2}{8}+\frac{B^2 \Lambda  \phi ^6}{8 A^2}+\frac{B K \phi ^6}{2 A^3}+\frac{K}{B^2} \ln \left(\frac{A}{A-B \phi ^2}\right)~.
\end{eqnarray}
The $f(R)$ model avoids the afforementioned tachyonic instability when $f_{RR}>0$, and for the obtained $f(R)$ function  we have
\begin{equation}
f_{RR}= -\frac{A^2 r^2}{128 K (B+r)^2}>0 \Rightarrow K<0~.
\end{equation}
For a particular combination of the scalar charges: $B=A/8$, the $f(R)$ model is simplified and takes the form:
\begin{equation} f(R) =R -6 \Lambda + \frac{128 K}{A^2} \ln \left(1-\frac{A^2 (R-6 \Lambda )}{128 K}\right) \end{equation}
The metric function (\Ref{b_ph}) as we can see, is similar to the BTZ black hole with the addition of a $\mathcal{O}(r)$ term because of the presence of the scalar field, and this term gives Ricci scalar its dynamical behaviour. The potential satisfies the conditions
\begin{equation} V(r \rightarrow \infty) = V(\phi \rightarrow 0) =0~, \end{equation}
and also $V'(\phi =0) =0$. It has a mass term which is given by
\begin{equation} m^2 = V''(\phi=0) = -\frac{3}{4} \Lambda~. \end{equation}
The metric function for $\Lambda = -1/l^2$ (AdS spacetime) and for $A,B>0$ has a positive root, since $K<0$. For $\Lambda = 1/l^2$ (dS spacetime) the metric function is always negative provided for $A,B>0$ and $K<0$, therefore   we will discuss only the AdS case.

The horizon is located at
\begin{equation} r_{+} = \frac{2 l \left(\sqrt{K \left(4 K l^2-A B\right)}-2 K l\right)}{A}~, \end{equation}
where we have set $\Lambda = -1/l^2$.
As we can see, in this  $f(R)$ gravity theory we have a hairy black hole supported by a phantom scalar field.

\begin{figure}[H]
\centering
 \includegraphics[width=.40\textwidth]{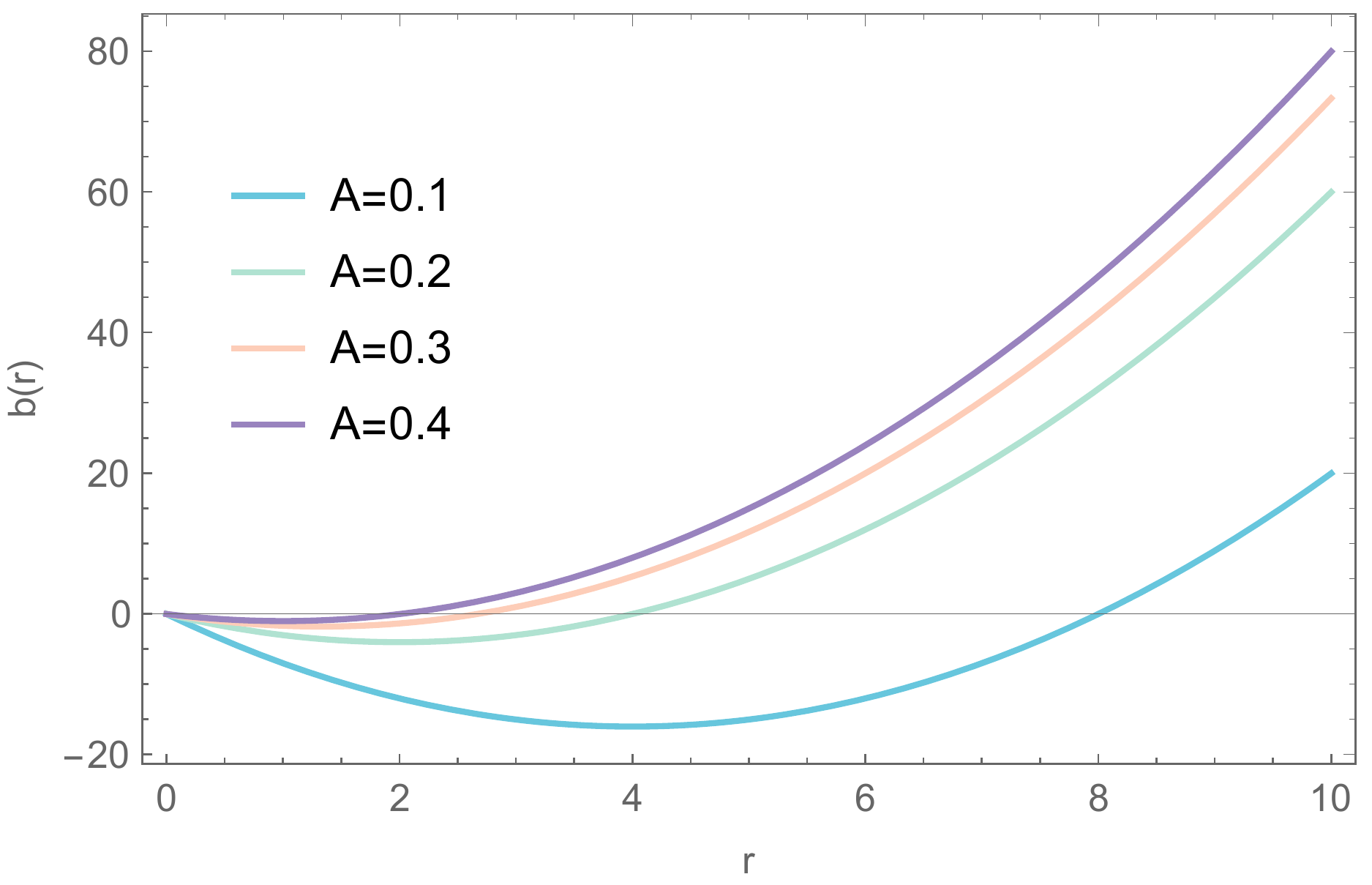}
 \includegraphics[width=.40\textwidth]{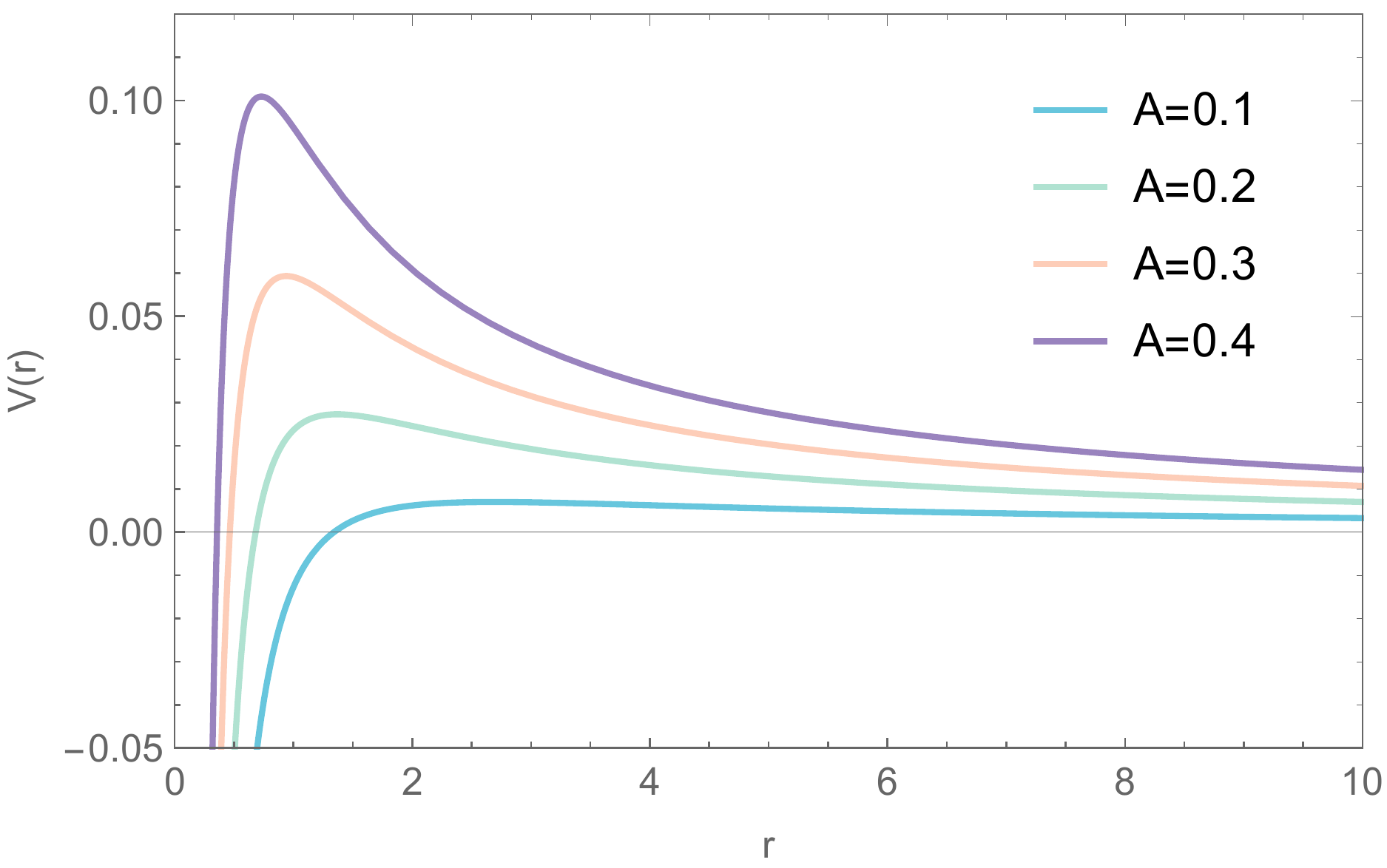}
 \includegraphics[width=.40\textwidth]{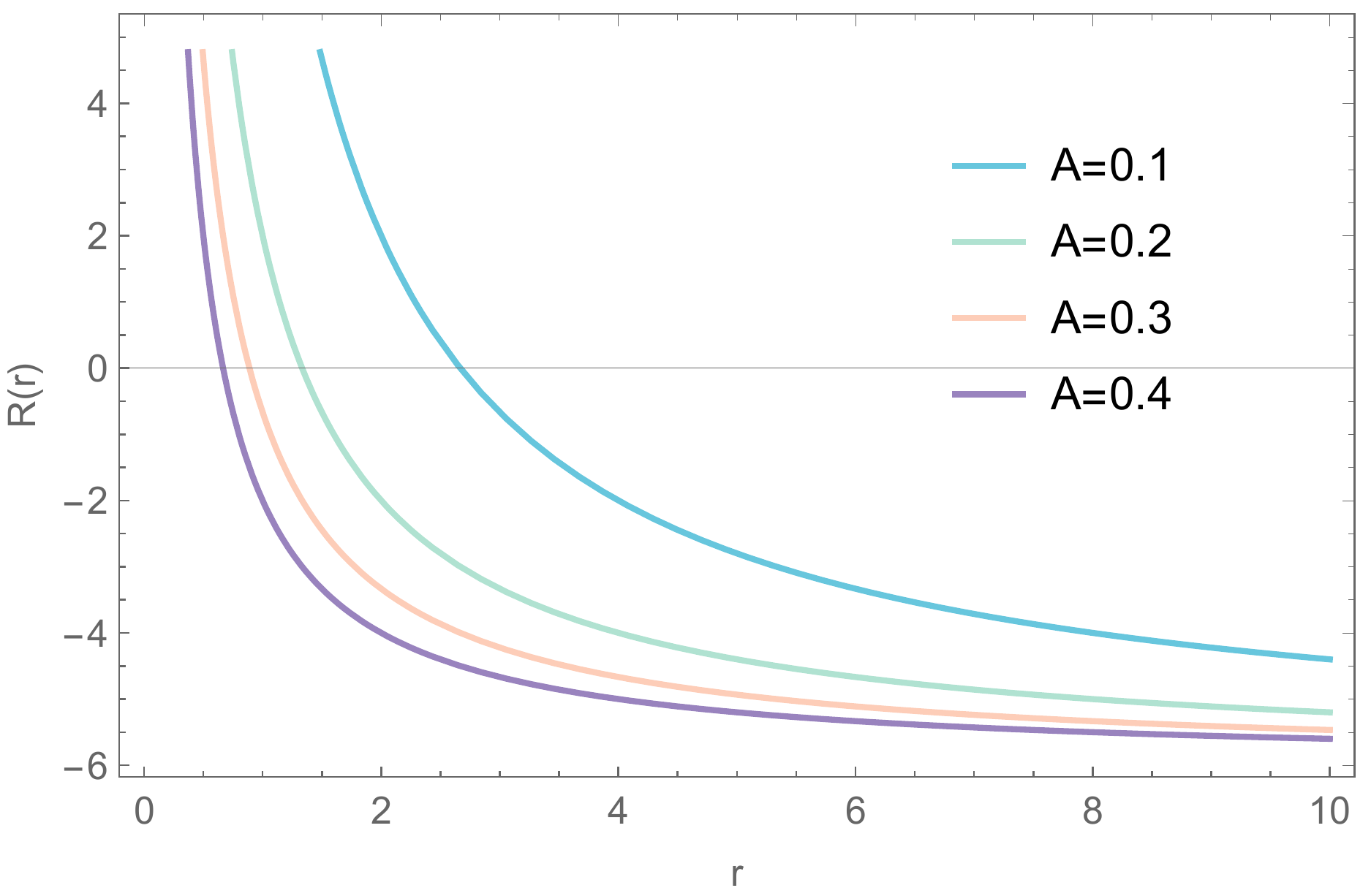}
 \includegraphics[width=.40\textwidth]{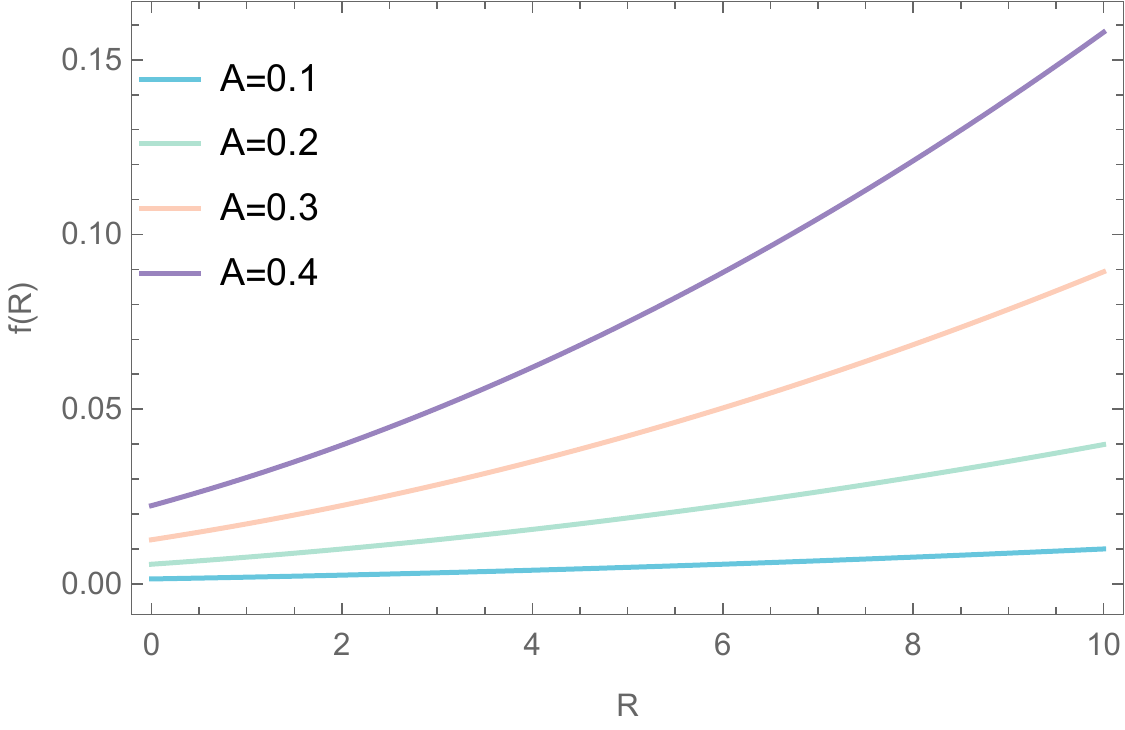}
\caption{We plot the metric function, the potential, the Ricci scalar and the $f(R)$ function of the phantom black hole for different scalar charge $A$, where other parameters have been fixed as $B=A/8$, $K=-1$ and $\Lambda=-1$. } \label{phantom_figure}
\end{figure}

\begin{figure}[H]
\centering
 \includegraphics[width=.40\textwidth]{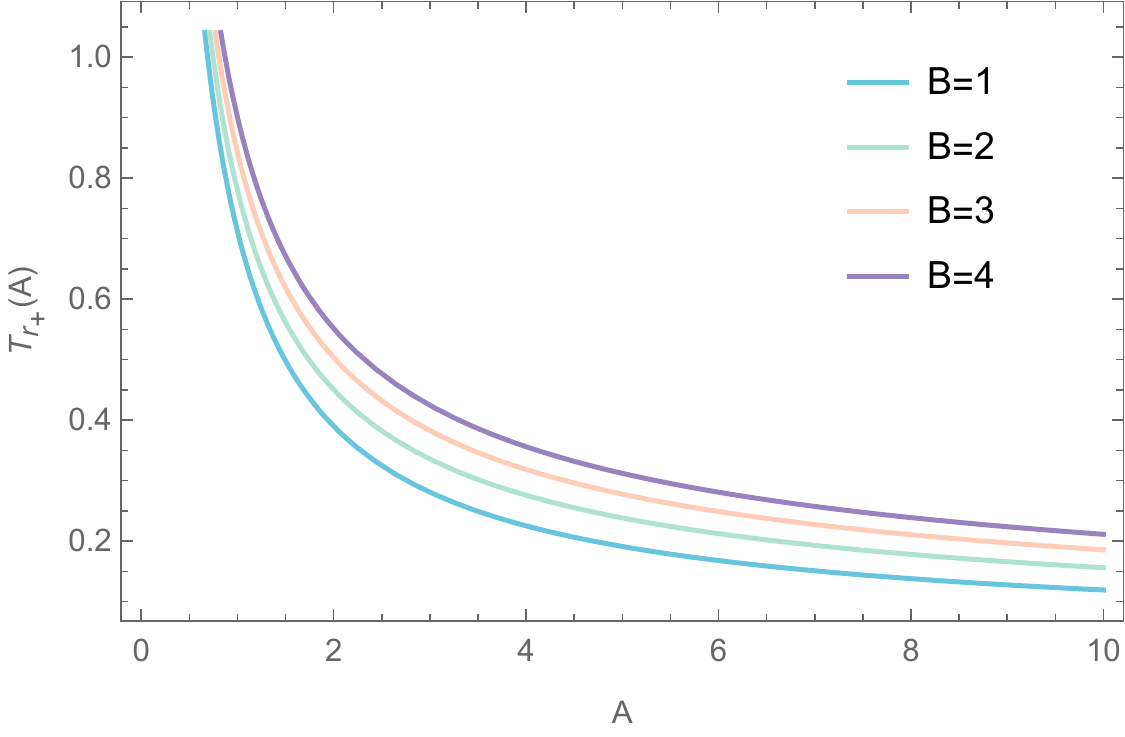}
 \includegraphics[width=.40\textwidth]{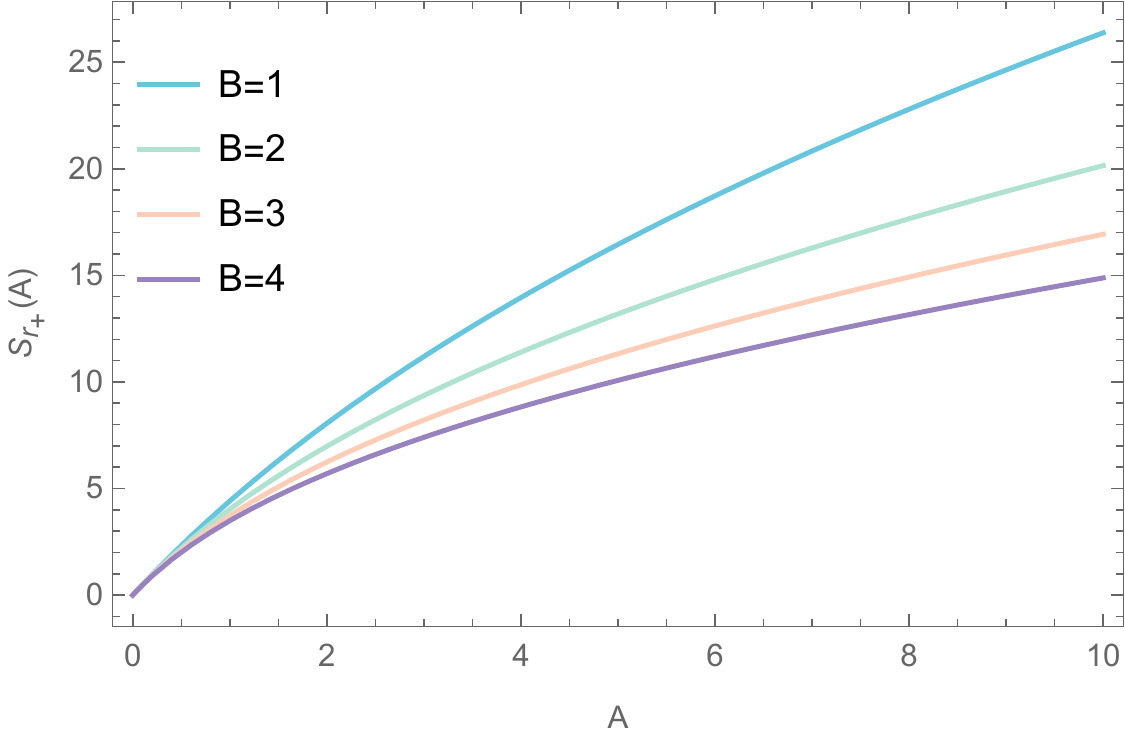}
\caption{The temperature and the entropy at the horizon of the black hole, as functions of the scalar charge $A$ while changing scalar charge $B$.} \label{phantom_thermodynamics}
\end{figure}

In FIG. \ref{phantom_figure} we show the behaviour of the metric function  $b(r)$, the potential $V(r)$, the dynamical Ricci scalar $R(r)$ and the $f(R)$
function. As can be seen in the case of $B=A/8$, the scalar charge $A$ plays an important role on the behaviour of the above functions. For example if the scalar charge $A$ is getting  smaller values the radius of the horizon of the black hole is getting larger. This means that even a small distribution of phantom matter can support a hairy black hole.

Looking at the thermodynamic properties of the model the Hawking temperature at the horizon is given by
\begin{equation} T(r_+) = \frac{2 K}{\pi  A}+\frac{r_+}{2 \pi  l^2} = \frac{\sqrt{K \left(4 K l^2-A B\right)}}{\pi  A l}~, \label{phantom_temp}\end{equation}
which is always positive for $A,B>0$ and $K<0$, while the Bekenstein-Hawking entropy is given by
\begin{equation}
S(r_+) =\frac{\mathcal{A}}{4G}f_{R}(r_+) =4\pi^2 r_+f_{R}(r_+)=\frac{A\pi^2 r_+}{2 (B+r_+)} = -\frac{\pi ^2 A K l}{\sqrt{K \left(4 K l^2-A B\right)}}>0~.
\label{entro}
\end{equation}

For the thermodynamic behaviour of the hairy black hole we can see from FIG. \ref{phantom_thermodynamics} that for larger  scalar charge $A$ we are getting smaller  temperatures, while the entropy has the opposite behaviour.

\section{Conclusions}
\label{sect4}

In this work, we considered $(2+1)$-dimensional $f(R)$ gravity with a self interacting scalar field as a matter field. Without specifying the form of the $f(R)$ function we derived the field equations and we showed  that the $f(R)$ model has a direct contribution from the scalar field. At first we considered the case, where $f_R(r) = 1 - \int \int \phi '(r)^2 drdr$, which indicates that if we integrate with respect to the Ricci scalar we will obtain a pure Einstein-Hilbert term and another term that depends on the scalar field. The asymptotic analysis of the metric function unveiled the physical meaning of our results. At infinity we obtain a scalarized BTZ black hole. The scalar charges appear in the effective cosmological constant that is generated from the equations. Corrections in the form of spacetime appear as $\mathcal{O}(r^{-n})$ (where $n \ge 1$) terms that depend purely on the scalar charges. At the origin we obtain a different solution from the BTZ black hole, where $\mathcal{O}(r)$ and $\mathcal{O}(r^2 \ln(r))$ terms change the form of spacetime.

 The scalar curvature is dynamical and due to its complexity it was difficult to obtain an exact form of the $f(R)$ function. Using asymptotic approximations, we show that the scalar charges make our theory to deviate form Einstein's Gravity. In the obtained results we considered the Dolgov-Kawasaki stability ctiterion \cite{Hendi:2014mba} to ensure that our theory avoids tachyonic instabilities \cite{DeFelice:2010aj,Bertolami:2009cd,Faraoni:2006sy}. We then calculated the Bekenstein-Hawking entropy and the Hawking temperature of the solution and we showed that the hairy solution is thermodynamically preferred since it has higher entropy.

 We then considered a pure $f(R)$ theory supported by the scalar field. We showed that thermodynamic and observational constraints require that the pure $f(R)$ theory should be builded with a phantom scalar field. The black hole solution we found has a metric function which is similar to the BTZ solution with the addition of a $\mathcal{O}(r)$ term. The scalar charge  is the one that determines the behaviour of the solution. For bigger scalar charge, the horizon radius is getting smaller meaning that the black hole is formed closer to the origin. The $\mathcal{O}(r)$ term is the one that gives to the Ricci scalar its dynamical behaviour. The obtained $f(R)$ model is free from tachyonic instabilities. We computed the Hawking temperature and the Bekenstein-Hawking entropy to find out that they are both positive, with the temperature getting smaller with the increase of the scalar charge  while the entropy behaves the opposite way, growing with the increase of the scalar charge.

 In the $f(R)$ gravity theories if a conformal transformation is applied from the original  Jordan frame to the Einstein frame then,  a new scalar field appears
which is  coupled minimally to the conformal metric and also a scalar potential is generated. The resulted theory can be considered as a scalar-tensor theory with  a geometric (gravitational) scalar field. Then it was shown in \cite{Canate:2015dda,Canate:2017bao}, that this geometric scalar field cannot dress a $f(R)$ black hole with  hair. On the other hand on cosmological grounds, it was shown in \cite{Capozziello:2014zda} that dark
energy can  be considered as a geometrical
fluid that adds to the conventional stress-energy tensor, which means that the determination of the 
dark energy equation of state depends on the understanding of which $f(R)$ theory  better fits current data. In our study we have 
introduced real matter parameterized by a scalar field coupled to gravity, therefore, it would be interesting to study the interplay of the geometric scalar field with the matter scalar field and see what are their implications to cosmology. However, to study this effect we have to extend  this work  to a study of $(3+1)$-dimensional $f(R)$  gravity theories. The main difficulty of constructing such theories is the complexity of their resulting equations. Nevertheless, even numerically we can get important information of how matter is coupled to $f(R)$ gravity and what are the  cosmological implications.

It would be interesting to extent this theory including an electromagnetic field. In three dimensions the electric charge makes  a contribution to the Ricci scalar, therefore we expect, like in the BTZ black hole, to find a charged hairy black hole in $f(R)$ gravity. One could also study the properties of the boundary CFT,  consider a rotationally symmetric metric anstaz to find rotating hairy black holes or study hairy axially symmetric solutions from hairy spherically symmetric solutions \cite{Capozziello:2009jg}.

 \acknowledgments
 We thank Christos Charmousis for very stimulating discussions.

\end{document}